\documentclass[aps,prb,twocolumn,groupedaddress,showpacs,floatfix,superscriptaddress,longbibliography]{revtex4-2}
\usepackage[plainpages=false,pdfpagelabels,colorlinks=true,linkcolor=red,urlcolor=blue,citecolor=blue,pdftitle={Title},pdfauthor={},pdfdisplaydoctitle=true,pdfduplex=DuplexFlipLongEdge]{hyperref}
\usepackage{epsfig}
\usepackage{amsmath,amssymb}
\usepackage{physics}
\usepackage{bm}
\usepackage{graphicx}
\usepackage[dvipsnames,usenames]{color}
\usepackage[normalem]{ulem}
\usepackage{soul} 

\definecolor{darkred}{rgb}{0.80,0,0}
\definecolor{blood}{rgb}{0.50,0,0}
\definecolor{brightred}{rgb}{1,0,0}
\definecolor{orange}{rgb}{1,0.3,0}
\definecolor{bluegreen}{rgb}{0,0.5,0.5}
\definecolor{lightblue}{rgb}{0,0.5,0.8}
\definecolor{darkgreen}{rgb}{0,0.5,0}
\definecolor{green}{rgb}{0,0.70,0}
\definecolor{darkblue}{rgb}{0,0,0.80}
\definecolor{magenta}{rgb}{1,0,1}
\definecolor{softmagenta}{rgb}{0.85,0.1,0.6}
\definecolor{mauve}{rgb}{0.6,0.1,1}
\definecolor{white}{rgb}{1,1,1}
\definecolor{black}{rgb}{0,0,0}
\graphicspath{{Images/}}

\begin{document}

\title{Quantum Membrane Phases in Synthetic Lattices of Cold Molecules or Rydberg Atoms}
\author{Chunhan Feng}
\affiliation{Department of Physics and Astronomy, University of California, Davis, CA 95616, USA}
\author{Hannah Manetsch}
\affiliation{Department of Applied Physics and Materials Science, California Institute of Technology,
Pasadena, CA 91125}
\author{Valery G. Rousseau}
\affiliation{5933 Laurel St., New Orleans, LA 70115, USA}
\author{Kaden R. A. Hazzard}
\affiliation{Department of Physics and Astronomy,
Rice University, Houston, TX 77005, USA}
\affiliation{Rice Center for Quantum Materials, Rice University, Houston, Texas 77005, USA}
\author{Richard Scalettar}
\affiliation{Department of Physics and Astronomy, University of California, Davis, CA 95616, USA}

\date{\today}

\begin{abstract}
We calculate properties of dipolar interacting ultracold molecules or Rydberg atoms in a semi-synthetic three-dimensional configuration -- one synthetic dimension plus a two-dimensional real space optical lattice or periodic microtrap array -- using the stochastic Green function Quantum Monte Carlo method.  Through a calculation of
thermodynamic quantities and appropriate correlation functions, along with
their finite size scalings, we show that there is a second order
transition to a low temperature phase in which two-dimensional `sheets' form in the
synthetic dimension of internal rotational or electronic  states of the molecules or Rydberg atoms, respectively.  Simulations for different
values of the interaction $V$, which acts between 
atoms or molecules that are adjacent both in real and synthetic space, allow us to compute a phase diagram. We find  a finite-temperature  transition at sufficiently large
$V$, as well as  a quantum phase transition -- a critical value $V_c$ below
which the transition temperature vanishes.
\end{abstract} 

\maketitle
\section{Introduction}

Atomic and molecular platforms for quantum simulation offer the versatility to realize an enormous range of physics, which has recently been extended to physics with extra dimensions by augmenting true  spatial extent with internal~\cite{boada12,celi:synthetic_2014,mancini:observation_2015,wall:synthetic_2016,sundar18,sundar19,kanungo21}  or motional states~\cite{Heimsoth_2013,gadway:atom_2015,price:synthetic_2017,an:nonlinear_2021}   coupled to mimic motion through a lattice in a synthetic dimension. 
Synthetic dimensions can lend access to many-body Hamiltonians with high levels of control over parameters such as
tunneling amplitude and phase in the synthetic dimension,  particularly useful 
for studying gauge fields, disorder, and topological band structures, all of which have been explored in recent experiments, reviewed in Ref.~\cite{ozawa19}. 

Recent experiments in ultracold molecules and Rydberg atoms demonstrate their potential as many-body systems with synthetic dimensions. Ultracold molecules' rotational-state-exchanging dipolar interactions  have been observed  in optical lattices~\cite{yan:realizing_2013,hazzard:many-body_2014,seesselberg:extending_2018}, they have been trapped  in  arrays of optical tweezers \cite{anderegg19,zhang2021optical}, and sophisticated coherent rotational
state  control of many levels has been demonstrated \cite{gregory19}, all the ingredients required to create many-body strongly interacting synthetic dimensions with rotational states of molecules
\cite{sundar18,sundar19}. Similarly, rapid advances in optical tweezer arrays  of dozens or hundreds of Rydberg atoms~\cite{scholl:quantum_2021, bluvstein:controlling_2021,Ebadi2021, semeghini:probing_2021,scholl2021microwaveengineering}
 can be combined with the synthetic dimensions for single Rydberg atoms demonstrated in Ref. \cite{kanungo21} to create many-body synthetic dimensions. Ref. \cite{kanungo21} specifically used a ladder of alternating $s$ and $p$ Rydberg states, but many alternative level schemes are possible. 
 
\begin{figure}[h!]
\centering
\includegraphics[width=0.85\columnwidth]{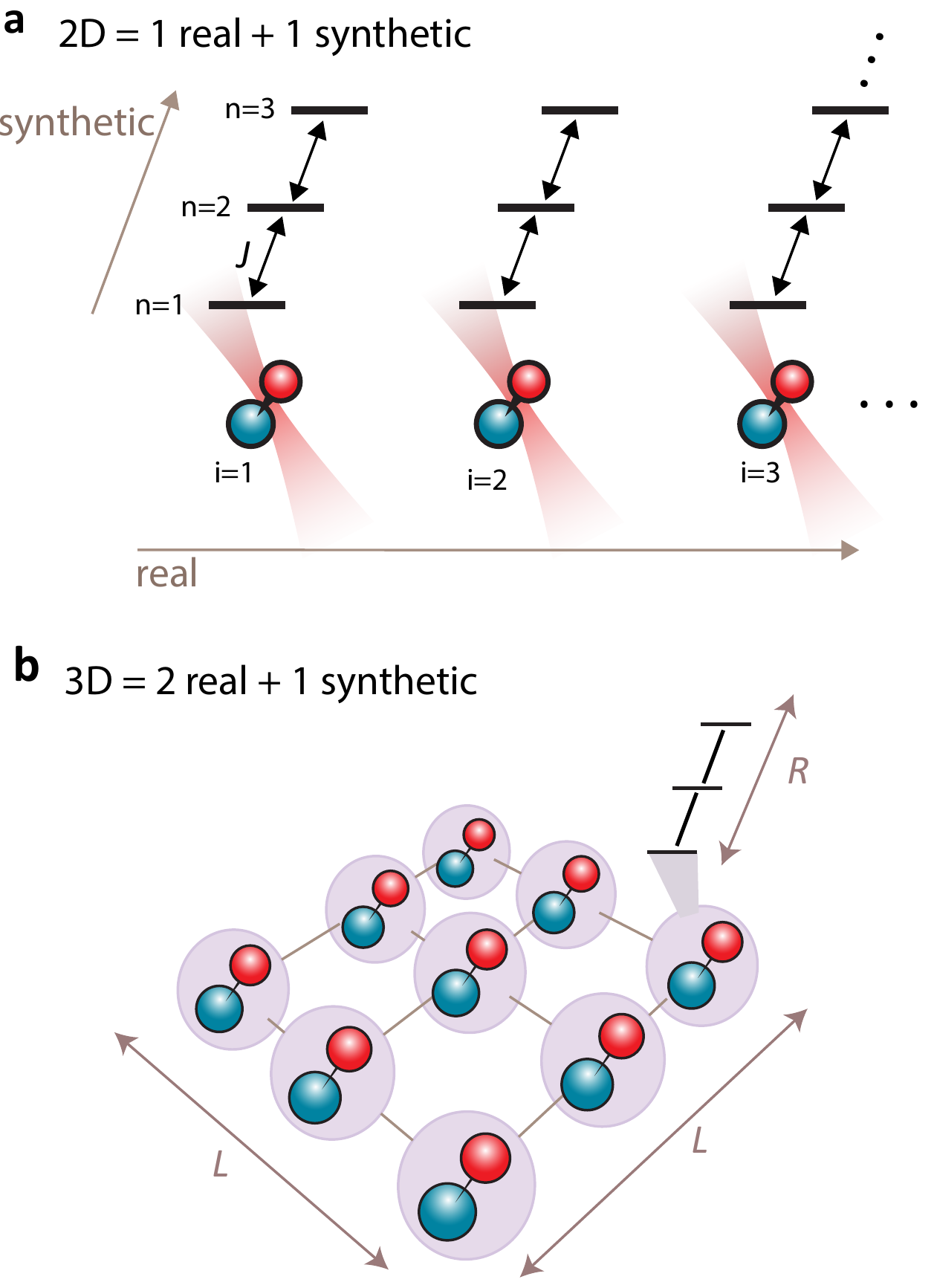}
\caption{{\bf (a)} A one-dimensional (1D) real and 1D synthetic-space lattice of ultracold molecules form an effective 2D system. Microwaves couple adjacent rotational states with synthetic tunneling rate $J$. {\bf (b)} 2D lattices augmented with a  synthetic dimension form an effective 3D system, the focus of this paper. 
}
\label{fig:synthetic-cartoon}
\end{figure}
 
Here, we consider an effectively three dimensional system constructed from a two-dimensional real-space lattice with one
 ultracold polar molecule or Rydberg atom per site, extended by a third, synthetic,
dimension of molecular rotational states, as illustrated in Fig.~\ref{fig:synthetic-cartoon}. We will mostly describe the system for the molecular case, but identical physics will apply for configurations of Rydberg atoms, as discussed below. We show that the system undergoes a quantum phase transition as  the dipolar interaction $V$, which causes angular momentum exchange between adjacent molecules, is varied relative to the ``tunneling" $J$ within the synthetic dimension. 

The transition represents the
development of a ``quantum membrane", or a low-temperature phase in which
molecules across the lattice are confined to occupy one of a few spontaneously chosen adjacent sites in the synthetic dimension, a two-dimensional surface fluctuating in the three dimensional space. Previously, this
phase transition was examined using  mean-field theories
\cite{sundar18}, and, in 1D, with  density matrix renormalization group (DMRG) calculations and  analytical solutions in the limit  $U/J\to\infty$ \cite{sundar19}. Here, we  utilize the Stochastic Green
Function (SGF) method \cite{Rousseau08} to perform numerically exact Quantum Monte Carlo
(QMC) calculations determining the phase boundary at zero and finite-temperature, as well as multiple observables across the phase diagram.  We evaluate the average
distance between molecules along  the synthetic direction, which measures the
degree of rotational state confinement into a membrane, as well as thermodynamic
quantities such as energy, heat capacity, and entropy, and correlation functions.  Varying the 
lattice size  allows a characterization of the phase transition in the
thermodynamic limit.

\section{Model and Methodology}

Experiments on two dimensional periodic arrays of ultracold polar
molecules in an optical trap or microtrap array with no real-space tunneling and with near-resonant microwaves coupling a chain of rotational states are  described by~\cite{sundar18} the  many-body
Hamiltonian,
\begin{align}
\cal{\hat H} = -&\sum_{i,n} J_n^{\phantom{\dagger}} \big(\, c_{i,n}^{\dagger} c_{i, n+1}^{\phantom{\dagger}}
+ c_{i,n+1}^{\dagger} c_{i, n}^{\phantom{\dagger}} \,\big)
\nonumber \\
+&\sum_{\langle i,j \rangle,n}
V_n^{i,j}
c_{i,n+1}^{\dagger} c_{i, n}^{\phantom{\dagger}} 
c_{j,n}^{\dagger} c_{j, n+1}^{\phantom{\dagger}}  \,\,.
\label{eq:quantumham} 
\end{align}
when the microwave transition rates, detuning and interactions are small compared to the energy differences between rotational states. Here  $c_{i,n}$ is either a bosonic or fermionic annihilation operator for a molecule at real space lattice site $i$ and rotational state (synthetic site) $n$, and the Hamiltonian is the  same  for both cases, since particles are frozen on their lattice sites and cannot exchange in real space. The first term arises from the near-resonant microwaves, where $J_n$ is proportional to the microwave amplitude coupling states $n$ and $n+1$. The second term arises from dipolar interactions, which  drive an exchange of angular momentum $n$ between pairs of molecules with adjacent angular momenta $(n_1, n_2)$, i.e. $(n,n+1) \rightarrow (n+1,n)$.  We have truncated the $1/r^3$ dipole interaction to nearest-neighbors. This captures the dominant interactions, and in 2D the longer range interactions are likely to give mostly quantitative corrections to the physics. More significant qualitative effects are expected only in small regions, for example at  low temperatures where multiple phases of matter with nearby energies compete.

Interpreting 
the rotational states as a \textit{synthetic dimension} indexed by $n$, the
first term describes microwave-induced tunneling along the synthetic dimension for each spatial location $i$ in the array,
while the second term describes the dipole interactions, which causes  molecules adjacent in the synthetic dimension to undergo coordinated quantum fluctuations.  The precise functional forms for
 $V_n^{i,j}$ are nontrivial \cite{sundar18}, but they vary slowly and approach a constant for large $n$, so we assume $V_n=V$ for all $n$, an excellent approximation for  many choices of rotational states. The $J_n$  are fully tunable in experiment by controlling the microwave amplitudes, but here we restrict to the 
simplest scenario, in which $J_n=J$ and  $V_n^{i,j}=V$.
Such a model provides the simplest example of an interacting synthetic dimension in this system, but as has been argued in 
Refs.~\cite{sundar18,sundar19} and we will see here, the physics is already quite rich, with interesting finite-temperature and quantum phase transitions.

In this work, we fix $J=1$ and $k_{\rm B}=1$, while focusing on  $V<0$, where the QMC sign problem is absent~\cite{loh90}. We refer to $R$ and $L$
as the linear lattice sizes in synthetic and real space, as illustrated in Fig.~\ref{fig:synthetic-cartoon}(b).
For a system of trapped, ultracold molecules or Rydberg atoms, $R$ is tunable and controlled by the microwave configurations applied. Initial experiments have demonstrated $R=6$ for Rydberg atoms~\cite{kanungo21} and demonstrated the individual couplings required  for $R=6$ for molecules~\cite{blackmore:coherent_2020}. No issues are expected to arise scaling to much larger $R$, and interesting physics already occurs for $R\le 6$. 
$N=L \times L$ denotes the total number of
real sites (molecules) on a square lattice. We use periodic
boundary conditions (PBC) for both real space and the synthetic dimension. Both open boundary conditions (OBC) and PBC in synthetic dimensions are realizable experimentally and we do not expect the choice of boundary conditions to alter the physics for large $R$ or $L$.

The SGF method \cite{Rousseau08} that we  use is closely related to the
Canonical Worm (CW) algorithms\cite{evertz93,prokofev98,VanHoucke06, Rombouts06}.  
It has the advantage of ease of implementation for general
forms of inter-particle interaction, but is somewhat less efficient
than methods which have been optimized for particular models.
Its name derives from the central role of the
many-body Green functions and its ability to capture them
in a simple and
general way.  The SGF approach works in the canonical ensemble 
and utilizes continuous imaginary
time. Hence it avoids any systematic errors introduced by
Trotter discretization
\cite{trotter58,suzuki76,fye86}.

\section{Results}  
\label{sec:holstein_interface}
\subsection{Thermal Phase Transition} 

\begin{figure}[t]
\centering
\includegraphics[width=\columnwidth]{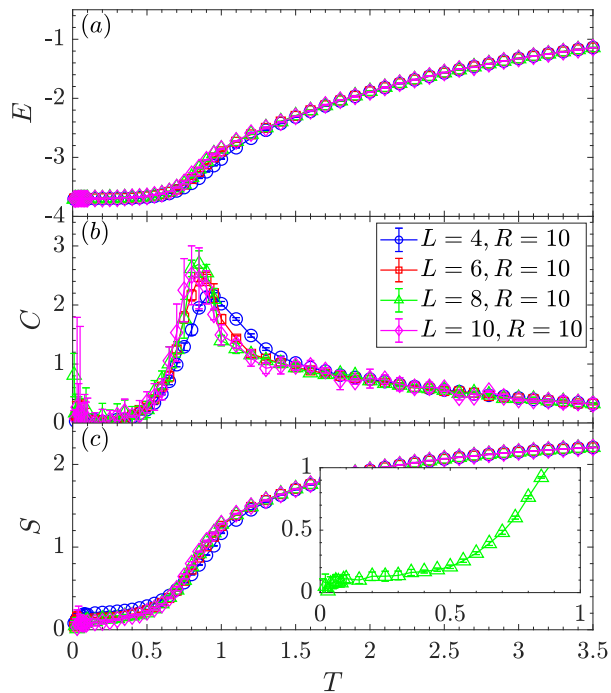}
\caption{{\bf (a)} Energy $E$, {\bf (b)} heat capacity $C$, and {\bf (c)} entropy $S$ as 
functions of temperature $T$ for $J=1$ and $V=-2$ for 
linear lattice sizes $L=4,6,8,10$ and  $R=10$ synthetic sites.  A signature of a phase transition is
present at $T \sim 0.8\, J$.  The inset to panel (c) highlights
a more subtle feature:  an entropy plateau separating
the phase transition from a final entropy drop
when the temperature is sufficiently small to
resolve a nearly degenerate set of states.
}
\label{fig:ECS_T_R10V2J1}
\end{figure}

We begin our determination of the phase diagram 
by considering a fixed number of rotational states $R=10$ and 
$V=-2\,J$.  The temperature dependence of the energy $E/N$ (i.e.~per spatial 
site), as obtained by the SGF, is shown in Fig.~\ref{fig:ECS_T_R10V2J1}(a) 
for  linear lattice sizes $L=4,6,8,10$. Figs.~\ref{fig:ECS_T_R10V2J1}(b,c) complete the picture of the
thermodynamics by showing the resulting heat capacity 
$C$, and entropy $S$. 
$C(T)= dE/dT$ 
is obtained from a numerical differentiation of the energy.
The entropy is obtained by   thermodynamic integration,
where
$T_{\text{max}} \sim 2\,|V|$ is the highest temperature we simulate. We have checked convergence in $T_{\text{max}}$ and find that it causes only a small shift of the curves, and doesn't change any of the features we identify.
We approximate $S(T_{\text{max}},V)$ by  $S(T_{\text{max}},V=0)$, the non-interacting entropy at $T_{\text{max}}$, which captures the zeroth order
term in the high-$T$ expansion. 

The most prominent feature of Fig.~\ref{fig:ECS_T_R10V2J1} is a sharp
decrease in both energy and entropy at $T \sim 0.8$, and the corresponding  peak in $C(T)$.  These are signals of a thermal phase transition, as
we will verify more rigorously below.  
In concurrence with this picture, the greater dependence of $E(T)$ 
on lattice size $L$   in the region $0.7 \lesssim T/J \lesssim 1.2$
 suggests  a large correlation length.

The inset of Fig.~\ref{fig:ECS_T_R10V2J1}(c) focuses on the low $T$ behavior of the entropy for
$L=8, R=10$, which reveals a plateau in $S(T)$ in a
temperature range $ 0.05\,J \lesssim T \lesssim 0.3\,J$
which can be understood
as follows:  In the limit $J=0$, the ground state is 
degenerate \cite{sundar19}.  
When $J$ is turned on, this ground state degeneracy is broken, so that the
resulting unique ground state yields an entropy $S \rightarrow 0$ as $T
\rightarrow 0$.  Our value $|J/V| = 1/2$, is sufficiently small that the
reduction in entropy occurs in two stages:  a phase transition to sheet
formation (as we shall show) at $T \sim 0.8\,J$, and then a final
decrease at
much lower $T$ when the temperature drops below the energy scale
distinguishing the nearly degenerate states, 
remnants of those exactly present at $J=0$.
There is a corresponding second peak in the heat capacity
Fig.~\ref{fig:ECS_T_R10V2J1}(b) at $T \sim 0.03\,J$.
\begin{figure}[t]
\centering
\includegraphics[width=\columnwidth]{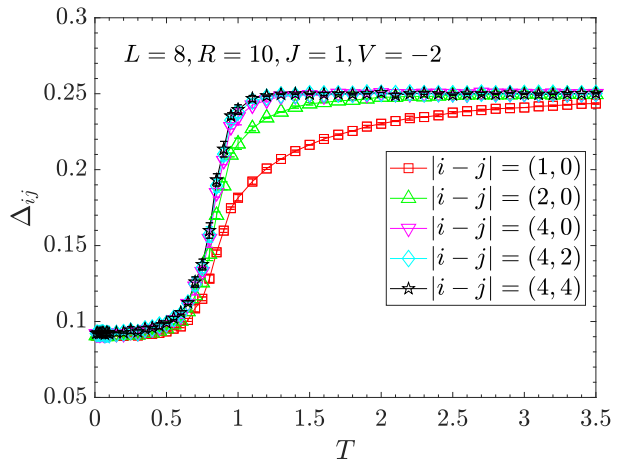}
\caption{Average distance $\Delta_{ij}$ (Eq.~\ref{eq:deltaij})
between the rotational
states of molecules on sites $i$ and $j$, as a function of temperature
$T$,
for dimensions $L=8, R=10$ and energy scales $J=1, V=-2$.
A drop, which becomes more abrupt as $|i-j|$ increases,
indicates long range binding of rotational states on different molecules.
}
\label{fig:deltaij_T_L8R10V2J1}
\end{figure}

In order to explore the nature of the ordered phase
suggested by the thermodynamics, we measure $\Delta_{ij}$, the average 
separation in the synthetic dimension  
between two molecules on spatial sites $i$ and $j$.
This is
\begin{align} 
\Delta_{ij} \equiv \frac{1}{R}\sum_{m,n} |m-n| \langle
c^\dagger_{i,m}c^{\phantom{\dagger}}_{i,m}
c^\dagger_{j,n}c^{\phantom{\dagger}}_{j,n} \rangle \,\,,
\label{eq:deltaij}
\end{align}
where $|m-n|$ denotes the distance between synthetic sites $m$ and $n$, accounting for periodic boundary conditions.

Measuring $\Delta_{ij}$ and related observables discussed below is challenging, but may be possible, in near-term experiments. Experimentally,  one can  
 measure populations in specific synthetic sites using standard techniques. In molecules, one can measure populations by state-selective absorption imaging~\cite{wang:direct_2010} or by absorption imaging the atoms produced after a STImulated Raman Adiabatic Passage (STIRAP) process addressing specific  rotational states~\cite{ni:high_2008,chotia:long-lived_2012}. However, this can be done only for a single rotational level per experimental shot, because the absorption imaging is destructive. Thus one can obtain only the $m=n$ terms in the integrand of Eq.~\eqref{eq:deltaij}. These drawbacks may be overcome by novel imaging techniques, such as  dispersive imaging~\cite{guan:nondestructive_2020}   or perhaps by taking advantage of  special molecules with quasi-cycling transitions~\cite{cheuk:Lambda_2018}. In Rydberg atoms, one can measure populations by selective field ionization~\cite{gallagher:rydberg_1994}  or level-specific transitions followed by absorption imaging, but novel non-destructive high-resolution real-space imaging will be required to directly measure correlations. 

Figure~\ref{fig:deltaij_T_L8R10V2J1} shows $\Delta_{ij}$ as a function of temperature for several $i,j$ for an $R=10$, $L=8$, $V=-2J$ system. $|i-j|$ denotes the vector connecting real sites $i$ and $j$. At high temperature, molecules
are randomly distributed and densities on
different real sites are independent. In this case, defining $P_m(i) \equiv c^\dagger_{i,m} c^{\phantom{\dagger}}_{i,m}$,  
we have
$\langle P_m(i)P_n(j) \rangle 
=\langle P_m(i)\rangle
\langle P_n(j)\rangle
=1/R^2=0.01$ when $i\neq j$.
As a
consequence, we find  $\Delta_{ij} \sim 0.25$ in the high temperature
limit when the sum over $m$ and $n$ is performed, in agreement with the high-$T$ limit of
Fig.~\ref{fig:deltaij_T_L8R10V2J1}. 
The transition to the 
relatively low value $\Delta_{ij} \sim 0.08$ below
$T_c \sim 0.8$, is consistent with the peak position
of $C(T)$, and signals the formation of an ordered quantum membrane phase,
with greatly reduced  separation in the synthetic dimension. 

\begin{figure}[t]
\centering
\includegraphics[height=2.5in,width=3.5in]{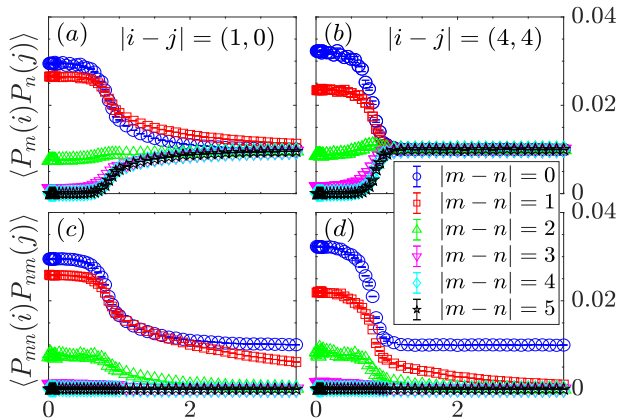}
\caption{$\langle P_m(i)P_n(j) \rangle $ and $\langle P_{mn}(i)P_{nm}(j\rangle )$ as functions of
temperature $T$ for a $L=8, R=10$ system with $J=1, V=-2$.
As $T$ is reduced below $T \sim 0.8$ the probabilities that the particles are in nearby synthetic states, $ |m-n| =0, 1$ rapidly grow, consistent with
entry into a membrane phase.
}
\label{fig:PmiPnj_T_L8R10V2J1}
\end{figure}

We can more fully characterize this phase by measuring
two additional correlation functions. The first, 
$\langle P_m(i)P_n(j)\rangle$,
breaks the average distance $\Delta_{ij}$ into its components and better resolves the rotational separation.
When 
$\langle P_m(i)P_n(j)\rangle$ 
is measured for the {\it same} molecule, $i=j$,
it takes either of the two temperature independent values $1/R = 0.1$ for
$n=m$, or $1/R^2 = 0.01$ for $n \neq m$.
On the other hand, for distinct molecules, $i \neq j$,
there is a strong temperature dependence.
At high $T$, $\langle P_m(i)P_n(j) \rangle \rightarrow 1/R^2 $ 
for all $n,m$, as discussed above, consistent with a complete lack of
correlation between the rotational states.
However, as $T$ is lowered, there is a rapid increase
in the probability that the rotational states are identical, 
$|m-n| =0$,
or adjacent,
$|m-n| =1$.
Simultaneously, $\langle P_m(i)P_n(j)  \rangle$ becomes small for all 
$|m-n| >2$.
This is true both for neighboring molecules 
$|i-j|=(1,0)$ and for those at maximal separation
$|i-j|=(4,4)$ for $L=8$, indicating the ordering of molecules to nearby synthetic locations is
long-ranged in real space.  

The results of Fig.~\ref{fig:deltaij_T_L8R10V2J1}  together with the top row of Fig.~\ref{fig:PmiPnj_T_L8R10V2J1}   give considerable insight into the nature of the phase transition and the ordered phase. At low-temperature, the string is mostly confined  to three adjacent states  (spontaneously chosen) in the synthetic dimension, and when the system is heated to the phase transition, the width of the strings diverges until it saturates the full width of the synthetic dimension. In an infinite system, it is expected to diverge at the phase transition, as investigated below. Intriguingly, the low-temperature width of the strings is reminiscent of the $J=0$ limit of the 2D=(1 real)+(1 synthetic) system at $T=0$ that was exactly solved in Ref.~\cite{sundar19}, which had a width of exactly three synthetic sites. Our results suggest the same phenomena occurs in 3D=(2 real)+(1 synthetic dimension), and that adding $J$ broadens the strings by a finite amount until the phase transition is reached and the string width diverges.

\begin{figure}[hbpt]
\includegraphics[width=1\columnwidth]{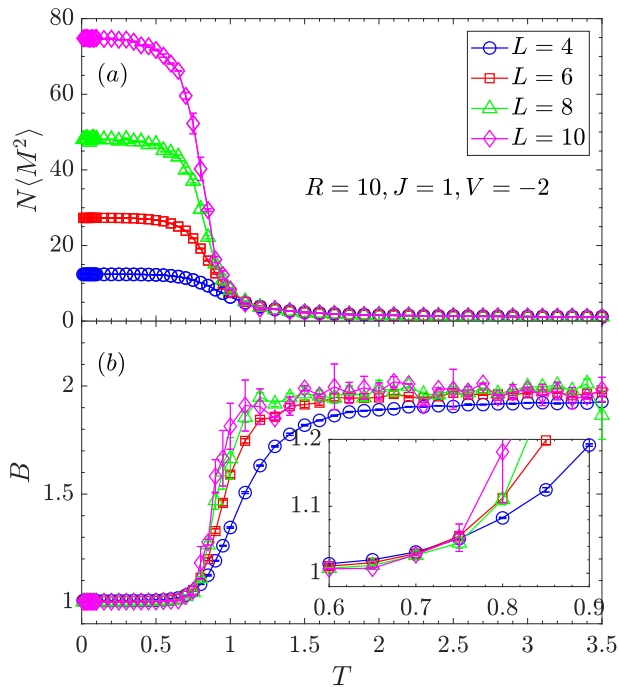}
\caption{{\bf (a)} Structure factor $N\langle M^2 \rangle$ as a function of
temperature $T$ when $J=1, V=-2$ for  
$L=4,6,8,10, R=10$. {\bf (b)} Binder ratio $B=\langle M^4\rangle/\langle
M^2\rangle ^2$ as a function of $T$. Inset: zoom-in showing  a crossing at $T_c \sim 0.75$, which coincides with
Fig.~\ref{fig:ECS_T_R10V2J1}-\ref{fig:PmiPnj_T_L8R10V2J1}.
}
\label{fig:M2_T_binder}
\end{figure}

We also consider the  correlation function, 
$\langle P_{mn}(i)P_{nm}(j)\rangle$ with $P_{mn}\equiv c^\dagger_{i,m} c^{\phantom{\dagger}}_{i,n}$, 
which
describes the probability of molecules 
$i,j$ being in states with superpositions of exchanged synthetic positions  $m$, $n$. 
As with
$\langle P_m(i)P_n(j) \rangle$, the correlation $\langle P_{mn}(i)P_{nm}(j)\rangle$
takes a trivial value $1/R$ for $|i-j|=(0,0)$.
For $|i-j|=(1,0)$ and $|m-n| =1$, this rotational exchange
is the expectation value of the second (interaction) term of 
the Hamiltonian (with the coefficient $V$ removed).  
Fig.~\ref{fig:PmiPnj_T_L8R10V2J1} (bottom) shows this correlation rapidly grows as temperature is lowered below $T \sim 0.8$.
The increase is even more abrupt for larger
$|i-j|=(4,4)$.
This corroborates the preference for
the occupation of nearby rotational states across the entire $8 \times 8$
molecular array for $T\lesssim 0.8$.

We now quantify these transitions through a finite size scaling analysis and an appropriately defined order parameter.
We begin,
in analogy to the common analysis of the Potts model\cite{Wu1982}, by defining an order parameter
\begin{align}
 M \equiv \sum_{m=1}^{R} e^{2 \pi i m/R} \frac{\sum_j P_j(m)}{N} 
\end{align}
that gives a one-dimensional  representation of the  group of translations in the synthetic dimension.
Then $N\langle M^2 \rangle$, shown in Fig.~\ref{fig:M2_T_binder}(a) can be
viewed as the corresponding structure factor. Above $T_c$, the synthetic position is random, and thus preserves synthetic translation invariance so $\langle M^2 \rangle
\sim \frac{1}{N} \rightarrow 0$ in the thermodynamic limit.  Below
$T_c$, in the quantum membrane phase, the structure factor $N\langle M^2
\rangle$ becomes proportional to the real-space lattice size $N$. 

An associated Binder ratio \cite{Binder1981} is
$B=\langle M^4\rangle \, / \, \langle M^2\rangle ^2$.  
Its crossings can be used to determine precisely both thermal and
quantum critical points. In the molecular gas phase, above
$T_c$, $\langle M^4 \rangle \sim 2/N^2$ and the Binder ratio $B
\sim 2$.  In the quantum membrane phase, below $T_c$, $\langle M^4
\rangle \sim \langle M^2 \rangle ^2$ and $B \sim 1$.  Precisely this
behavior is seen in Fig.~\ref{fig:M2_T_binder}(b).
A  Binder ratio crossing 
of curves for different lattice sizes occurs, and is
emphasized  in the inset.
This allows an accurate determination of $T_c = 0.75 \pm 0.02$, refining the more crude
$T_c$ estimates from
Figs.~\ref{fig:ECS_T_R10V2J1}-\ref{fig:PmiPnj_T_L8R10V2J1}. 

\begin{figure}[hbpt]
\includegraphics[width=1\columnwidth]{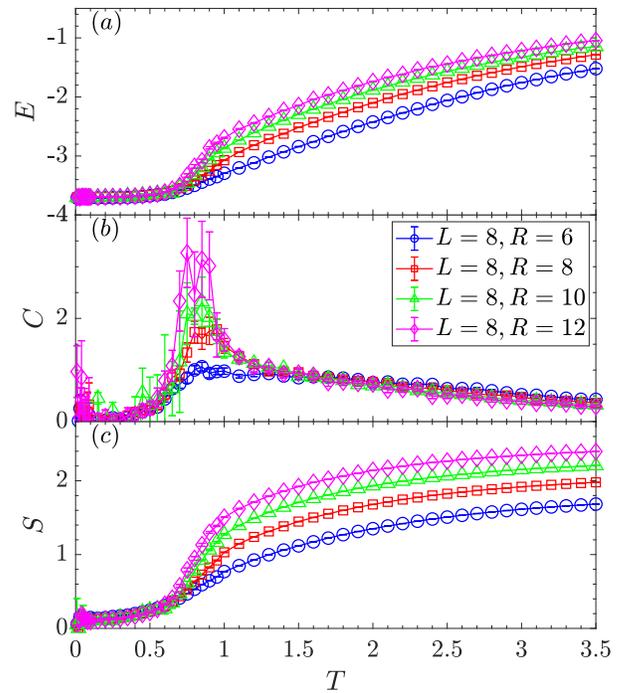}
\caption{{\bf (a)} Energy $E$,  {\bf (b)} heat capacity $C$, and {\bf (c)} entropy $S$ as a
function of temperature $T$ for $J=1,V=-2$, and
$R=6,8,10,12$.}
\label{fig:ECS_T_L8V2J1_diffR}
\end{figure}

We  now expand our focus from the $R=10$ results considered so far to study
the dependence of the thermal phase transition on $R$, shown in
Fig.~\ref{fig:ECS_T_L8V2J1_diffR}. When the temperature $T$ is lower than
$T_c$ (which can depend on $R$), molecules collapse to a quantum membrane with thickness of roughly 2 or 3 sites
and the total energy is independent of the number of unoccupied synthetic sites, as is
 seen in 
Fig.~\ref{fig:ECS_T_L8V2J1_diffR}(a). 
For $T>T_c$, the energy $E$  depends on $R$, increasing with $R$ as the molecules  escape the bound states. The heat capacity peaks increase with
$R$, reflecting the greater loss in entropy as the membrane forms 
and the occupied rotational states decline from $\sim R$ to $\sim 2$-$3$.
The positions of the peaks, i.e.~the locations of $T_c$, decrease as $R$ increases, but the
shift is relatively small.  The critical temperature of the square lattice Potts model, a rough classical
analog of the quantum model considered here, is known to decline as
$T_c \sim 1/{\rm ln}\big(\, 1 + \sqrt{R}\,\big)$, which similarly has a very
slow dependence on $R$ \footnote{Comparing $R=6$ and $R=12$,
$1/{\rm ln}\big(\, 1 + \sqrt{6}\,\big) = 0.808$ and
$1/{\rm ln}\big(\, 1 + \sqrt{12}\,\big) = 0.669$.
}

\begin{figure}[hbpt]
\includegraphics[width=1\columnwidth]{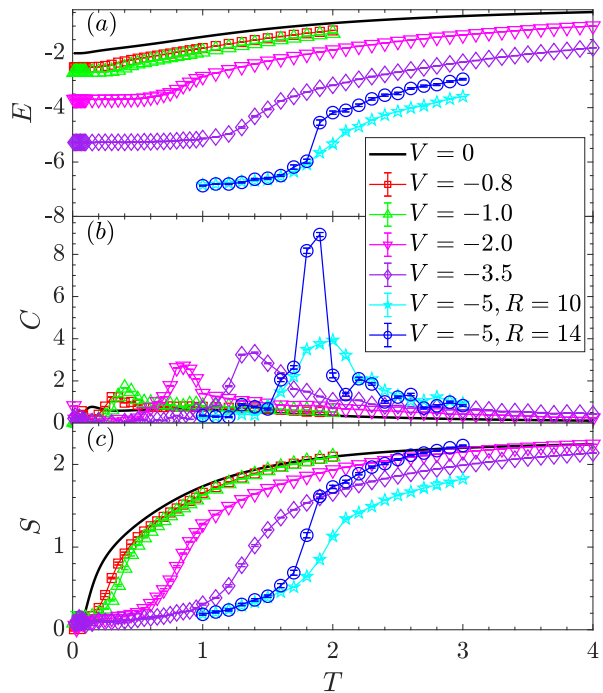}
\caption{{\bf (a)} Energy, {\bf (b)} heat capacity $C$, and {\bf (c)} entropy $S$ as functions of
temperature $T$ for $J=1$, $R=10$, $L=8$,  for  interaction
strengths $V=-5, -3.5, -2, -1, -0.8, 0$, and for $R=14, V=-5$.  
}
\label{fig:ECS_L8R10J1_diffV}
\end{figure}

We similarly expand our calculations from the previous  $V=-2$ to general $V$.
Analogous to Fig.~\ref{fig:ECS_T_R10V2J1}, the energy $E$, heat
capacity $C$  and entropy $S$ are shown in
Fig.~\ref{fig:ECS_L8R10J1_diffV}.   All the curves are for a $8 \times 8$
square lattice with $R=10$. The black curve is obtained from analytical
solution for the non-interacting limit $V=0$. As $V$ increases, the
transition temperature $T_c$ rises and for
$\left|V\right| \gtrsim 5$, $T_c$ is roughly proportional to $V$, an expected result
since $J$ is negligible and $V$ is the only energy scale at these temperatures. In the non-interacting limit, the ground
state degeneracy is completely removed and entropy drops to $S=0$ directly
without any plateaus. As the interaction strength $V$ grows, wider 
low temperature entropy plateaus are evident.

\subsection{Quantum Phase Transition} 

\begin{figure}[hbpt]
\includegraphics[width=1\columnwidth]{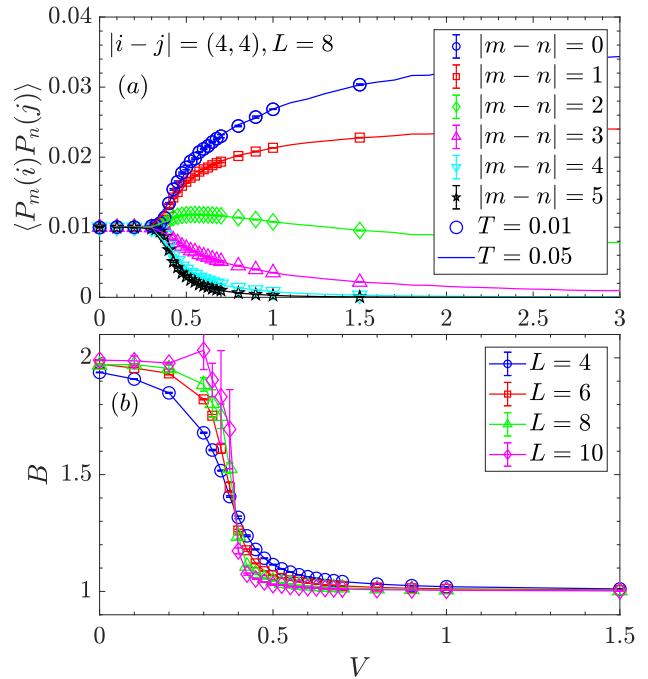}

\caption{{\bf (a)} $\langle P_m(i)P_n(j) \rangle$ as a function of $V$ for a $L=8$ and $R=10$ system
at low temperatures $T=0.05,0.01$. {\bf (b)}  The Binder ratio $B=\langle
M^4\rangle/\langle M^2\rangle ^2$ crossing shows a quantum critical
point  at $V_c = 0.40 \pm 0.025$. 
}
\label{fig:PmiPnj_binder_V_lowT}
\end{figure}

The results of  Fig.~\ref{fig:ECS_L8R10J1_diffV} show that $T_c$ is becoming small as $V$ decreases. 
This suggests the possibility that there is a quantum critical point (QCP) below which there is
no ordered phase (no membrane formation) even at zero temperature.
To explore this possibility, we fix $T$ at a low value 
and vary the interaction strength $V$.
The correlation function 
$\langle P_m(i)P_n(j) \rangle$ is plotted in
Fig.~\ref{fig:PmiPnj_binder_V_lowT}(a). Both $T=0.05$ (solid line) and
$T=0.01$ (dot) curves are presented.  The fact that they coincide
indicates we have accessed the ground state for the given finite lattice.
The data support the existence of a QCP
at $V_c \sim -0.4$.  Below this value, even in the ground state,
$\langle P_m(i)P_n(j)\rangle = 1/R^2$ for all $|m-n|$, its uncorrelated,   gas phase, value.
However, for $|V| > 0.4$, the quantum membrane
phase occurs at low $T$. 
$\langle P_m(i)P_n(j) \rangle$ is then large for $|m-n| \leq 2$, whereas
$\langle P_m(i)P_n(j)\rangle \sim 0$ for $|m-n| > 2$ away from $V_c$. 
The values of 
$\langle P_m(i)P_n(j) \rangle$ for $|m-n| > 2$ grow as $V_c$ is approached: 
the membrane becomes increasingly thick.

Similar to  the thermal phase transition, 
a more accurate way to
locate $V_c$ is by plotting the Binder ratio $B$ as a function of $V$ for
different lattice sizes.  The 
clear crossing of Fig.~\ref{fig:PmiPnj_binder_V_lowT}(b) determines  
the position of the QCP, $V_c = 0.40 \pm 0.025$.

\subsection{Phase Diagram} 
\begin{figure}[hbpt]

\includegraphics[width=1\columnwidth]{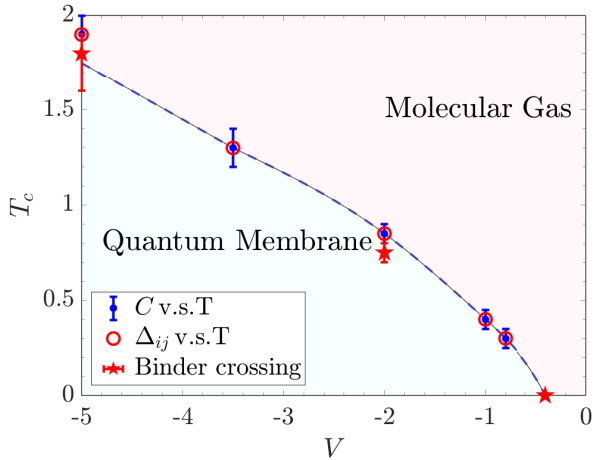}
\caption{
Phase diagram of the Hamiltonian Eq.~\ref{eq:quantumham}
describing ultracold polar molecules in two spatial and one synthetic dimension.
Transition temperature $T_c$ (obtained from the peak position in heat capacity curves, the largest slope position in $\Delta_{ij}$ vs. $T$ plots and the location of the Binder crossing) as a function of $V$ for $R=10$ system.}
\label{fig:Tc_V}
\end{figure}

Putting together the results for the thermal and quantum phase transitions  leads to
the phase diagram shown in Fig.~\ref{fig:Tc_V}. The blue dots are
obtained from the peaks in the heat capacity curves $C(T)$ and the red open circles are
extracted from the temperature of largest slope in the average distance
$d\Delta_{ij}/dT$.  Both data sets were analyzed at fixed spatial lattice size, an $8 \times 8 $
square with $R=10$ rotational states.  They are in close quantitative agreement. As is suggested in Fig.~\ref{fig:ECS_T_L8V2J1_diffR}, the dependence of transition temperature $T_c$ on number of rotational states $R$ near this value $R=10$ is weak.

As noted earlier, the most reliable $T_c$ are obtained from a Binder crossing,
as in Fig.~\ref{fig:M2_T_binder} for $V=-2$.
These are shown by red stars at $V=-2$ and $V=-5$ in Fig.~\ref{fig:Tc_V}.
The Binder value lies somewhat below the values inferred from a single lattice size,
an expected result. 
Finally, the
red star at $T_c=0$ reveals the quantum critical point implied by the Binder
crossing in Fig.~\ref{fig:PmiPnj_binder_V_lowT}.  The two phases are labeled
in the figure:  When $T>T_c$, the
system is in a molecular gas phase where spatially separated molecules are uncorrelated. When $T<T_c$ there is a quantum
membrane phase where molecules separated far in real space remain close along the synthetic direction, with $|m-n|$ a finite value, approaching $\lesssim 2$ as $V\to  -\infty$.

\section{Discussion and conclusions}

In this work we calculated properties of 3D systems formed by 2D real-space arrays of either ultracold polar molecules or Rydberg atoms augmented by  a synthetic dimension employing  the Stochastic Green Function (SGF)  quantum Monte Carlo (QMC) method.  Thermodynamic quantities -- the energy $E$,
heat capacity $C$, entropy $S$ were measured as
functions of temperature $T$ and interaction strength $V$ and suggested the occurrence of
both finite temperature and quantum  phase transitions. The high-temperature 3D gas transitions to a low-temperature quantum membrane, a  spontaneously formed fluctuating 2D surface, as evidenced by  correlation functions that capture the average separation in the synthetic dimension [$\Delta_{ij}$], probabilities for two molecules to occupy given synthetic sites [$\langle P_m(i)P_n(j) \rangle$],
and a measure of correlated hopping in the synthetic dimension for particles separated in real space [$\langle P_{mn}(i)P_{nm}(j) \rangle$].
Finite size studies of these observables, including 
an analysis of the Binder ratio, suggest a genuine phase transition and capture $T_c$ accurately.   Together
with the extraction of 
a quantum critical coupling
strength $V_c$, these data produced
a phase diagram in the plane of temperature $T$ and the energy scale $V$ of
coordinated quantum tunneling of adjacent molecules.

We show that the quantum membrane phase suggested in prior work~\cite{sundar18,sundar19}  zero-temperature mean-field or in 2D (1 real+1 synthetic) DMRG calculations also exists in 3D (2 real+1 synthetic), and that it survives to finite temperature. We also show this  with significantly larger systems and finite-size scaling, and with a variety of conceptually important and experimentally accessible observables. These elucidate the nature of the thermal and phase transitions as a continuous transition occurring by the divergence of the string width.
These results will be invaluable for experiments that are beginning to probe synthetic dimensions in arrays of trapped molecules and Rydberg atoms. 

A question opened by these results is the precise nature of the classical and quantum phase transitions, in particular their universality class. It is unclear whether this corresponds to a known  universality class, or something previously unexplored. Additionally, while the results should capture the main physics, the long-ranged $1/r^3$ interactions will lead to quantitative, and perhaps some qualitative, modifications of the behavior, and will need to be incorporated for detailed comparison with future experiments.
Finally, the results and methods also serve as a jumping off point for calculations in the infinite variety of synthetic landscapes that can be experimentally engineered, including topological band structures, such as the Su-Schrieffer-Heeger model realized in Ref.~\cite{kanungo21}, models with disorder, ladders, gauge fields, or  even exotic geometries such as M\"obius strips. The interplay of the interaction-driven tendency towards a quantum membrane with the single-particle band structures is expected to lead to a  rich variety of  physics. 

\vskip0.10in
\noindent
\underbar{\bf Acknowledgements:} We thank Sohail Dasgupta and Bryce Gadway for useful conversations.  
H.M.~was supported by the Research Experience for Undergraduates
program (NSF grant PHY-1852581) and by the Caltech Applied Physics Department Yariv/Blauvelt Fellowship.
The work of C.H.F.~and R.T.S.~was
supported by the grant DOE-DE‐SC0014671 funded by the U.S. Department of
Energy, Office of Science. K.H.~was supported by the Welch Foundation
Grant No.~C1872, 
the National Science Foundation  Grant
No.~PHY1848304, and also  benefited from discussions at
the KITP, which was supported in part by the National Science Foundation
under Grant No.~NSF PHY-1748958.

\bibliography{synthetic}

\end{document}